\documentclass[%
 reprint,
 amsmath,amssymb,
 aip,
]{revtex4-2}
\usepackage{color}
\usepackage{graphicx}
\usepackage{dcolumn}
\usepackage{bm}
\usepackage[hyperindex,breaklinks]{hyperref}

\begin{document}

\preprint{APS/123-QED}

\title{Machine Learning for Phase Behavior in Active Matter Systems}

\author{Austin R. Dulaney}
 \affiliation{Division of Chemistry and Chemical Engineering, California Institute of Technology}


\author{John F. Brady}
 \email{jfbrady@caltech.edu}
\affiliation{Division of Chemistry and Chemical Engineering, California Institute of Technology}%

\date{\today}

\begin{abstract}
We demonstrate that deep learning techniques can be used to predict motility induced phase separation (MIPS) in suspensions of active Brownian particles (ABPs) by creating a notion of phase at the particle level. 
Using a fully connected network in conjunction with a graph neural network we use individual particle features to predict to which phase a particle belongs.
From this, we are able to compute the fraction of dilute particles to determine if the system is in the homogeneous dilute, dense, or coexistence region.
Our predictions are compared against the MIPS binodal computed from simulation. 
The strong agreement between the two suggests that machine learning provides an effective way to determine the phase behavior of ABPs and could prove useful for determining more complex phase diagrams.
\end{abstract}

\maketitle


\section{\label{sec:intro}Introduction}

Since its inception, the field of active matter has been dominated by studies of motility-induced phase separation (MIPS). 
The majority of these studies focus on developing a theoretical framework to describe clustering behavior and the accumulation of active particles at boundaries.
Due to the striking similarities between classical and active phase behavior, the creation of thermodynamic-like frameworks has been of particular interest but continues to be a source of debate. \cite{Fily2012AthermalAlignment, Takatori2015a, Levis2017ActiveCoexistence, Klamser2018ThermodynamicMatter, Solon2018GeneralizedEnsembles} 
A key difficulty surrounding this approach is the lack of a well-defined notion of temperature and free energy---as these systems are far from equilibrium---which results from the intrinsic swimming motion of active particles.

Adhering to the structure of traditional thermodynamic frameworks has resulted in several definitions of a non-equilibrium chemical potential,\cite{Takatori2014, Chakraborti2016AdditivityParticles, Paliwal2018ChemicalState} each of which predicts an active binodal but fails to predict the correct coexistence pressure measured inside the phase envelope from simulation. 
The shortcomings with the current chemical potential definitions do not preclude its existence but necessitate alternative measures for determining the phase boundaries. 
Large scale computer simulations provide a means to computing system pressure, which can provide insights into the phase behavior through the mechanical instability criterion. 
While this method is robust and has shown great success,\cite{Takatori2015a, Fily2018MechanicalMatter, Patch2018Curvature-dependentPhases} it inherently has a steep trade-off between accuracy and computational cost. 
Determination of the phase boundary requires the change in system pressure with volume fraction to be zero. 
To make such a judgment one either needs to finely sweep volume fraction space or rely on fitting functions to smoothly fit the pressure data. 
Both methods are highly dependent on the quality of the pressure data obtained at each point in phase space, and large fluctuations in active pressure make this a difficult task, especially deep in the coexistence region. \cite{Takatori2015a, Fily2018MechanicalMatter}

To overcome these limitations we turn towards methods used to characterize other inherently complex materials. 
Recently, there has been a surge of interest in using machine learning algorithms to aid in material characterization.\cite{Carrasquilla2017MachineMatter, vanNieuwenburg2017LearningConfusion,Suchsland2018ParameterNetworks, Swanson2020DeepMaterials}
While early studies were predominantly interested in materials containing explicit symmetries or those confined to two-dimensional lattices,\cite{Carrasquilla2017MachineMatter,vanNieuwenburg2017LearningConfusion} there has been some development in classifying amorphous materials.\cite{Swanson2020DeepMaterials} 

In this study, we leverage the developments in machine learning to aid in characterizing the observed phase behavior in suspensions of active Brownian particles (ABPs).
ABPs are an important minimal model system for determining the behavior of self-propelled colloids, bacteria, and other living organisms.
The key feature that distinguishes an active colloid from a passive one is the driven and persistent nature of its motion. 
This distinct characteristic of its dynamics gives rise to a wealth of interesting behaviors including self-assembly,\cite{Mallory2018} clustering,\cite{Palacci2013LivingSurfers, Bechinger2016} and motility-induced phase separation.\cite{Fily2012AthermalAlignment, Takatori2015a, Levis2017ActiveCoexistence, Klamser2018ThermodynamicMatter, Solon2018GeneralizedEnsembles}
As such, active materials have garnered interest from the chemical and material science communities for novel drug delivery methods, remediation strategies, and material design methods at the microscale.\cite{Mallory2018, Gao2014, Ebbens2016ActiveApplications}

Due to the nonequilibrium nature of these systems, it is difficult to develop analytic theories that can accurately predict the more complex collective behaviors. 
Thus, we look towards machine learning to aid in this endeavor.
Machine learning algorithms are capable of discerning difficult---and potentially nonintuitive---nonlinear relationships among system variables, which would otherwise go unnoticed. 
These algorithms also have the benefit of readily handling multi-body correlations, which are exceptionally taxing or intractable through traditional analytic means. 

Using a combination of deep learning and large-scale simulation, we focus on characterizing the phase behavior of particles in a suspension of active Brownian disks. 
We use machine learning to predict particle phase at a per particle level for simulations conducted at different regions in phase space. 
We then use these phase labels to get an estimate for the fraction of particles in each phase present at a given point in phase space.
This fraction is then used to determine the system phase behavior.
The manuscript is outlined as follows. 
In section~\ref{subsec:simulations} we define the implementation of the active Brownian particle model used in our simulations. 
We then outline the datasets generated for use in our machine learning model in section~\ref{subsec:datasets}.
Here we also discuss the feature selection used for our machine learning model.
In section~\ref{subsec:ml_framework}, we describe the machine learning model architecture used in this work and provide details on the training procedures.
In section~\ref{subsec:ml_feats}, we discuss the input features used for our model.
In section~\ref{sec:graph}, we discuss the representation of our simulation snapshots as graphs.
In section~\ref{sec:results}, we present our results in the form of predictions of the phase behavior for suspensions of ABPs at different regions in the phase diagram. 
Finally, in section~\ref{sec:conclusion} we discuss the implications of this work and future directions.

\section{\label{sec:methods} Methods}
\subsection{\label{subsec:simulations}Simulation Details}

Suspensions of monodispersed, purely active particles are modeled using the active Brownian particle (ABP) model. 
The active motion is characterized by an intrinsic swim velocity $\mathbf{U_{0}}=U_{0}\mathbf{q}$---where $\mathbf{q}$ is the particle orientation---which reorients on a timescale $\tau_{R}$. 
Particles of radius $a$ interact through a Weeks-Chandler-Anderson (WCA) potential with cutoff radius $r_{cut} = (2a)2^{1/6}$ and depth ${\epsilon = 200F^{swim}a}$, where ${F^{swim} = \zeta U_{0}}$ is the magnitude of the force resulting from the product of the translational drag $\zeta$ and swim velocity. 
Here we assume particles reorient via a stochastic torque $\mathbf{L}^{R}$ governed by zero-mean white noise statistics with variance ${2\zeta^{2}_{R}\delta(t)/\tau_{R}}$, where $\zeta_{R}$ is the rotational drag coefficient. 
Particle positions and orientations can be evolved in time using overdamped Langevin dynamics
\begin{align}
    0 =& -\zeta\mathbf{U}_{i} + \mathbf{F}^{swim}_i + \sum_{i \neq j}\mathbf{F}^{P}_{ij}, 
    \label{eq:langevin_trans}
    \\
    0 =& -\zeta_{R}\mathbf{\Omega}_{i} + \mathbf{L}_{i}^{R},
    \label{eq:langevin:rot}
\end{align}
\noindent where $\mathbf{F}^{swim}_{i} = \zeta U_{0} \mathbf{q}_i$ is the swim force of particle $i$, ${\mathbf{F}^{P}_{ij}}$ is the interparticle force between pair $i,j$, $\mathbf{U}_{i}$ is the velocity, $\mathbf{\Omega}_{i}$ is the angular velocity, and $\zeta$ and $\zeta_{R}$ are the translational and rotational drags, respectively.
The angular velocity is related to the particle orientation by $\partial \mathbf{q}_{i}/ \partial t = \mathbf{\Omega}_{i}\times \mathbf{q}_{i}$.
Normalizing position and time by $a$ and $\tau_{R}$, respectively, results in the dimensionless reorientation P\'eclet number $Pe_{R} \equiv a/l$, which is the ratio of a particle's size to its persistence length ${l = U_{0}\tau_{R}}$---the distance traveled between reorientation events.\cite{Takatori2015a}

We performed independent simulations of 40,000 particles for 10,000$\tau_{R}$ for various combinations of the two governing nondimensional parameters: the packing fraction $\phi$ and $Pe_{R}$.
To avoid introducing an additional force scale $Pe_R$ was varied by changing $\tau_{R}$ at a fixed value of $U_{0}$.
All simulations were conducted using the HOOMD-Blue software package.\cite{Anderson2008, Glaser2015}
Hydrodynamic interactions have been neglected.

\subsection{\label{subsec:datasets} Datasets}

Our machine learning model is structured to predict phase identity at a per particle level, similar to what was done by Ha \textit{et al}.\cite{Ha2018}
This results in a binary classification task in which particles can be members of the gas phase or the dense phase.
For simplicity, we ignore the second-order hexatic transition present in two-dimensional hard disk systems and treat the hexatic phase as part of the dense phase. 

We use the simulations outlined in section~\ref{subsec:simulations} to produce datasets for each point in phase space represented by a $(\phi,Pe_{R})$ pair. 
For each of these phase points, we look at 6 snapshots spaced roughly 1,000$\tau_{R}$ apart.
From these snapshots, we construct a feature set for each phase point which consists of 240,000 entries. 
Predictions of the phase behavior at each phase point are averaged across each of the 6 time points to reduce bias from a single configuration.

\subsection{\label{subsec:ml_framework} Learning Framework}
Here we give a brief overview of neural networks and describe the architecture and training routine used in this work.

Neural networks have shown great potential for predicting particle phase for both two-state and amorphous phase-separated systems.\cite{Suchsland2018ParameterNetworks, Swanson2020DeepMaterials} 
The most common neural network employed is the fully connected feedforward network. 
Feedforward networks are composed of layers of transformations modified by nonlinear functions. 
These layers can be stacked resulting in the output of one layer acting as the input of the following layer. 
The basic form for a layer $f$ is $f(x) = g(Wx + b)$, where $g$ is the nonlinear activation function, $x$ is the vector input data, $W$ is the weight matrix, and $b$ is a vector of biases. 
When constructing a fully connected network the activation functions $g$ for each layer need not be the same, and additional regularization terms can be added to prevent overfitting to training data. 
Some common activation functions are the sigmoid, hyperbolic tangent, and rectified linear unit (ReLU), defined as ${g(x) = max(0,x)}$. 
Once constructed, a network is given an objective, or loss, function to minimize and updates the weight and bias terms through either gradient descent~\footnote{Gradient descent is an iterative method for locating the local minimum of a function by determining the steepest gradient of the function with respect to its independent variables. In this context gradient descent is used with backpropagation, which relates the weights of each layer in the neural network back to the loss function so the entire network can be updated during each iteration.} or a more sophisticated algorithm like Adam.\cite{Kingma2015Adam:Optimization}
Here we are interested in a binary classification and thus use binary cross entropy to compute loss
\begin{equation}
    L = -(y \log (p) + (1-y)\log(1-p),
\end{equation}

\noindent where $L$ is the loss, $y$ is the binary indicator of whether the positive class is the correct label for a given observation, and $p$ is the probability that an observation is of the positive class. 

Recent advances in machine learning have resulted in the adoption of graph convolutional neural networks (GNNs), which utilize graph theory to add information on the spatial proximity of training data.\cite{Kipf2016Semi-SupervisedNetworks,Velickovic2018GraphNetworks,Wu2019ANetworks} 
These are similar to traditional convolutional neural networks (CNNs), which rely on convolution and connected layers to make predictions. 
The primary uses for CNNs have been been in the areas of computer vision and natural language processing, due to the inherent structure of image and text data. 
Similarly, GNNs use convolutions and the inherent structure of the data, but adjacent training points need not be distributed on a rectilinear grid like an image or sequentially like in text.\cite{Wu2019ANetworks} 
In both architectures, the input matrix is convolved with a set of matrices---the convolution layer---to produce output matrices. 
These convolution layers are equivariant under translation and rotation, making them highly effective at learning abstract features of an image or graph while simultaneously reducing the number of parameters. 

The amorphous configurations found in particle-based phase-separated systems can benefit from traditional CNNs,\cite{Swanson2020DeepMaterials} but this requires spatial discretization of the system which may vary with particles of different sizes. 
We avoid this when looking at MIPS in active disks by using a GNN to provide information on the local structure to the network. 

Our training and model architecture is as follows.
We first train a supervised deep neural network (DNN) on data in the single-phase region above the critical point. 
After the supervised network is trained, we predict particle labels for a simulation of interest.
These predictions are then taken and those that predict the phase with a  $>$90\% confidence are used as seed labels in a semi-supervised GNN. 
We then take the simulation snapshot and represent it as a graph, which we will discuss in more detail in section~\ref{sec:graph}.
We then train a GNN for each graph.
The training in this step is structured as a transductive, or semi-supervised, learning problem.
For each graph we use the seeded particle (node) labels to propagate labels to the remainder of the graph.
We use the same features from the DNN, but instead of learning a very general problem, we are using confidently labeled particles to influence the labels given to their neighbors.
In this work, we use the graph attention network (GAT) architecture\cite{Velickovic2018GraphNetworks} for our GNNs implemented using the DGL software package.\cite{Wang2019DeepNetworks}
The resulting prediction from the GNN is then weighted against the prediction provided by the DNN in the first step. 

\begin{figure}[tb]
    \centering
    \includegraphics{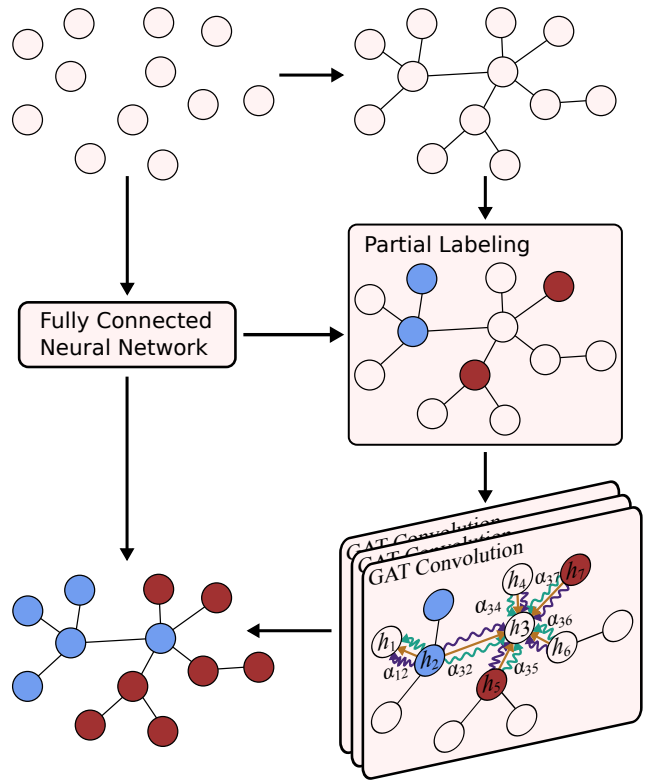}
    \caption{The learning architecture used in this work to predict particle phase labels. 
    First, a particle feature matrix is fed into a fully connected DNN.
    Simultaneously particles are connected to form a graph structure.
    The graph is partially labeled using the most confident ($>$90\%) labels from the DNN and is then used with the feature matrix as inputs into a GNN consisting of three GAT convolution layers with a final softmax activation function. 
    The resulting label probabilities from the GNN are then averaged with the label probabilities output from the DNN to achieve the final label probabilities.
    Each particle is then given the most probable label.}
    \label{fig:model_scheme}
\end{figure}

A flowchart of our learning process is outlined in Fig.~\ref{fig:model_scheme}.
The purple, orange, and teal lines of the GAT convolution represent the different attention heads for the layer. 
Each attention head serves as a means to create feature abstractions.\cite{Velickovic2018GraphNetworks}
The coefficients $\alpha_{i,j}$ are learned weight parameters which determine the weighted importance of neighbor $j$ on particle $i$.
The attention heads from each node are then concatenated or averaged to produce the layer output, which may be a label probability or feature abstraction.
Details of the GAT implementation can be found in reference \onlinecite{Velickovic2018GraphNetworks}.
Further details of the model architecture used in this work are presented in appendix~\ref{app:model_details}.

\subsection{\label{subsec:ml_feats} Feature Selection}

In order to label individual particles, our feature space is limited to quantities that can be computed on a per-particle basis. 
This includes Voronoi volume, the number of first shell Voronoi neighbors, and the average of first shell Voronoi volumes.
We repeat this averaging process for the second and third shell neighbors as well to incorporate information about the local environment.
We also include the hexatic and translational order parameters defined as $\psi_{6}(i) = 1/n \sum^{n}_{j}e^{\mathrm{i} 6\theta_{ij}}$ and ${ G_{6}(\mathbf{r}_{ij})=\sum^{n}_{j} \psi_{6}(i) \cdot \psi^{*}_{6}(j) }$, respectively, where $n$ is the number of Voronoi neighbors, $\mathbf{r}_{ij}$ is the vector connecting pair $ij$, $\theta_{ij}$ is the angle between $\mathbf{r}_{ij}$ and the reference vector ($0,1$), and $\psi_{6}^{*}(i)$ is the complex conjugate of the hexatic order parameter.
The Voronoi volumes and the hexatic and translational order parameters were computed using the Freud analysis software.\cite{Ramasubramani2019Freud:Data}
In order to account for some of the dynamics we include the force-orientation autocorrelation $\mathbf{F}_{i} \cdot \mathbf{q}_{i}$ and the particle speed $\mathbf{U}_{i} = \mathbf{F}/\zeta$.

\begin{figure*}[t]
    \centering
    \includegraphics[width=\textwidth]{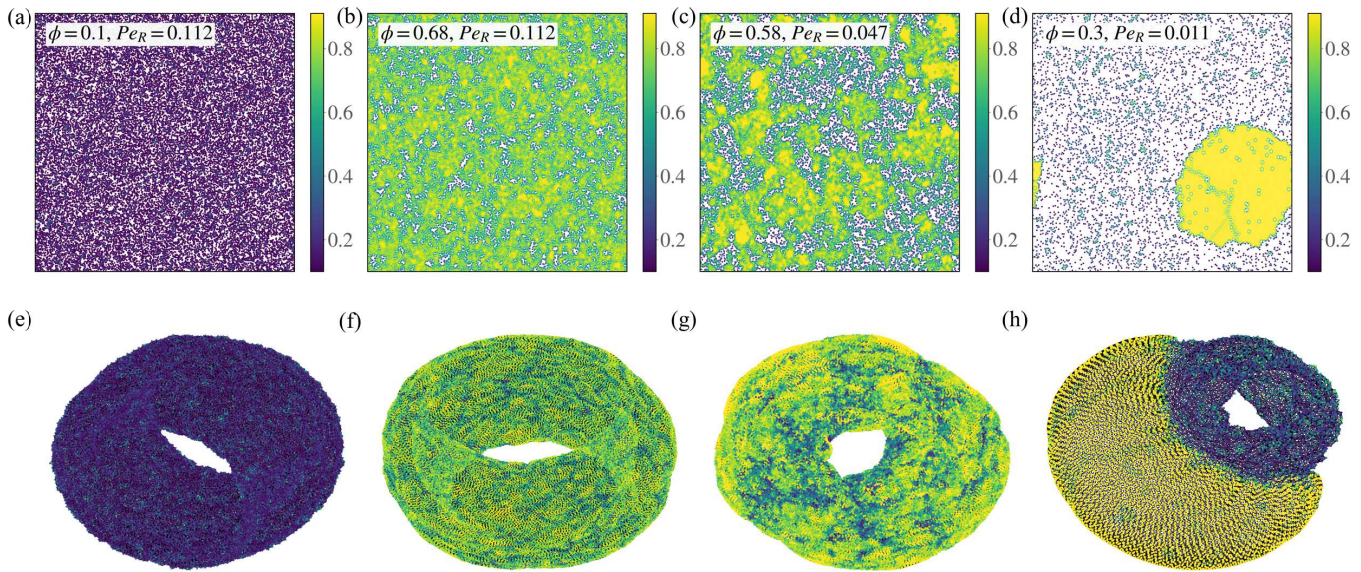}
    \caption{Simulation snapshots and respective graph structures for different regions of phase space colored by particle Voronoi volume.
    We look at the weakly active ($Pe_{R}\sim 0.11$) (a),(e) dilute and (b),(f) dense regions,  (c),(g) the region near the critical point ($P_{R} \sim 0.047$), and (d),(h) deep within the coexistence region ($Pe_{R}\sim 0.011$).
    }
    \label{fig:graphs}
\end{figure*}

Our initial set of features is paired down using a boosted random forest to remove highly collinear features in order of importance.
The final feature set is comprised of the Voronoi volume, number of third shell neighbors, hexatic order parameter, translational order parameter, and the force-orientation autocorrelation in order of importance. 
The process of removing collinear features is discussed further in appendix~\ref{app:feature_corr} and the correlation matrices for the full and final feature sets are shown in Fig.~\ref{fig:correlations}.
It is interesting to note that the number of third shell neighbors is ranked highly in importance because the model might be learning the order-disorder hexatic transition.
Lastly, we take each of our features and average them across all first shell neighbors to create an additional set of aggregate features. 
This aggregation step improved the performance and training stability of our DNN in the first step of our model.

\section{\label{sec:graph} Graph Representation}

The MIPS transition is markedly similar to the liquid-vapor transition seen in traditional thermodynamic fluids with the two coexisting phases both being disordered. 
In thermodynamic fluids one could measure local density and use spatial density discontinuities to distinguish between the coexisting phases.
However, this is difficult to do in practice as we are constrained to finite systems in simulations.
Regions close to the critical point are subject to large density fluctuations which make it difficult to observe persistent macroscopic phase domains. 
Therefore, we need an alternative way to gather this similar type of local structure in the system. 
We do this by representing the system as a graph.

For each simulation snapshot, we represent the system as a graph where each particle is a node in the graph and connections are made between first shell Voronoi neighbors. 
With periodic boundaries, this results in a fully-connected graph. 
If the edges are then constrained to be the distances between particles we obtain a three-dimensional, toroidal shape as depicted in Fig.~\ref{fig:graphs}(e)--(h).

In Fig.~\ref{fig:graphs}, we present simulation snapshots at different points of the phase diagram with their corresponding graph representations. 
Each particle and corresponding graph node are colored based on the Voronoi volume fraction of that particle.
The graphs in each region of the phase diagram possess unique morphologies and characteristics.
The gas phase [see Fig.~\ref{fig:graphs}(a),(e)] is marked by a uniform graph with a rough surface.
The disorder in the phase prevents a smooth surface from forming and any structure present is short-range.
If we next look at a primarily dense system [Fig.~\ref{fig:graphs}(b),(f)], we see that the graph representation still has bumps on the surface, but they are not as sharp.
The increased system density causes jamming and reduces the magnitude of fluctuations, which results in longer-range morphological features.
When we approach the critical point [see Fig.~\ref{fig:graphs}(c),(g)], the graph starts to form "lumps" which are connected by coarse sections of the surface. 
This results from mixing regions with dense and dilute phase features.
We can think of the connecting coarse regions as articulation points in the graph surface, which become less pronounced with lower activity.
As we go deep into the coexistence region, the dense region is made up of a single large crystal providing clear spatial domains for the two phases, as shown in Fig.~\ref{fig:graphs}(d).
The distinctive regions manifest as a coarse mesh for the dilute phase---similar to Fig.~\ref{fig:graphs}(e)---and a smooth surface for the dense phase with perturbations resulting from crystalline defects [see Fig.~\ref{fig:graphs}(h)].

The clear distinction in graph structure in the different phase regions lends support for the use of graph neural networks to aid in predicting particle phase. 
The use of local structure also serves to help make decisions for particles near phase interfaces and regions which may be marked by large density fluctuations.

\section{\label{sec:results}Results}

\begin{figure}[tb]
    \centering
    \includegraphics{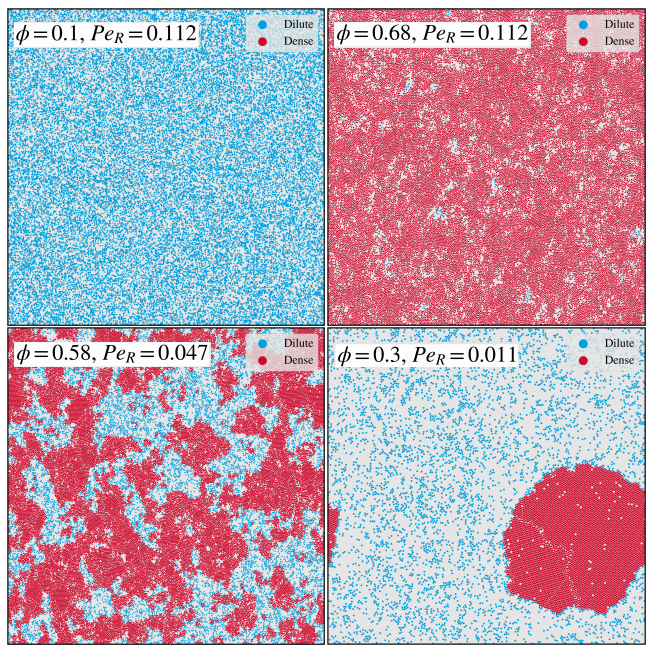}
    \caption{Simulation snapshots for different regions of the phase diagram with particles colored based on their predicted phase.}
    \label{fig:snapshots}
\end{figure}

Here we present the results from our machine learning model.
Our model was trained on very dilute ($\phi<0.2$) or very dense ($\phi > 0.7$) phase points above the critical point and deep within the coexistence region.
Training points from deep within the coexistence region were labeled by inspection and particles near phase interfaces were not used for training.
The model was then used to predict particle phase below the critical point ($Pe_R^{crit} \sim 0.047$).
Figure~\ref{fig:snapshots} presents the snapshots of particles shown in Fig.~\ref{fig:graphs}(a)--(d) colored by their predicted phase labels. 
It can be seen that our model is highly capable of distinguishing particle phase in the homogeneous phase and deep within the coexistence region.
The most challenging region is near the critical point as these particles are more difficult to readily distinguish from inspection.

Therefore, to evaluate performance in this region we take the predicted particle labels for each phase point and compute the fraction of dilute particles $F_{g}$ present and average this across all $T=6$ time points for a given ($Pe_{R}$,$\phi$) pair.
This is represented by
\begin{equation}
    F_g = \frac{1}{T}\sum_{t=0}^{T}\frac{1-\sum_{j}^{N}y_{j}}{N},
\end{equation}

\noindent where $y_j$ is the predicted label of particle $j$ and $N$ is the total number of particles. 
In our model the positive case is the dense phase ($y_{j}=1$) and the null case is the dilute phase ($y_{j}=0$).
To account for small fluctuations in prediction we consider a point to be in the dilute region if $F_g >$95\% and to be in the dense region if $F_{g}<$5\%.
Any other value of $F_{g}$ is labeled as coexisting as we are only considering points below the critical point ($Pe_{R}^{crit}\sim 0.0468$).

\begin{figure}
    \centering
    \includegraphics[width=0.49\textwidth]{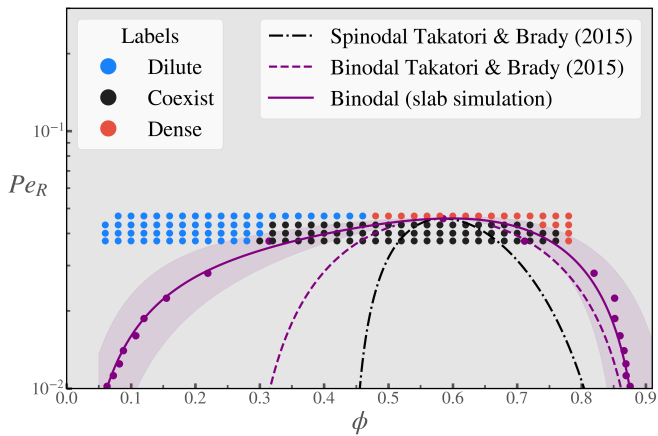}
    \caption{The $Pe_{R}$--$\phi$ phase diagram for purely active Brownian particles. 
    We show the spinodal (black dash-dotted line) and binodal (purple dashed line) predicted by Takatori and Brady\cite{Takatori2015a} along with the binodal computed from slab simulations (purple points).
    A fourth-order polynomial fit (solid purple line) is used to give a more complete picture of the computed binodal.
    The shaded region represents one standard deviation above and below the predicted fitting parameters.
    The remaining points on the graph are colored based on their predicted region of phase space from our machine learning model.
    We use a cutoff of $>$95\% gas fraction to be considered gas (purple) and $<$1\% gas fraction to be considered in the dense phase (blue).
    Every value for gas fraction between those values is considered within the coexistence envelope. 
    Here we show $Pe_{R}$ values in the range 0.0468--0.0374.
    }
    \label{fig:binodal}
\end{figure}

Figure~\ref{fig:binodal} presents the MIPS phase diagram with points colored based on which phase the system is predicted to be in using our machine learning model. 
We compare the predicted phase against the binodal predicted by Takatori and Brady\cite{Takatori2015a} (purple dashed line) and the binodal computed from slab simulations (purple points).
The solid purple line is a fourth-order polynomial fit of the computed binodal.
The shaded region represents the range of this fit $\pm 1$ standard deviation for each fitting parameter.
The spinodal predicted by Takatori and Brady (black dash-dotted line) is shown for completeness.
We find remarkable agreement between the binodal obtained from simulations and our machine learning predictions.

From Fig.~\ref{fig:binodal} it is clear that predicting dilute particles is more challenging than predicting dense particles. 
We suspect this difficulty arises from the large tail in the distribution of Voronoi densities for particles in the dilute phase. 
Ha \textit{et al.} observed this type of large overlap for particle density distributions when studying the phase behavior of a Lennard-Jones fluid.\cite{Ha2018}

\section{\label{sec:conclusion}Conclusions}

We have created a machine learning model to predict the phase identity of individual active Brownian particles.
Our results indicate that single-particle parameters are sufficient for learning particle phase when some amount of structure is included in the system. 
We have also shown that the MIPS phase transition can be predicted using this machine learning model.
From our model optimization and feature analysis, we conclude that kinematic features---such as particle speed or the force-orientation correlation---are important for distinguishing the phases present in the MIPS transition (see appendix~\ref{app:feature_corr} and Fig.~\ref{fig:shap}), unlike the traditional liquid-vapor transition present in thermodynamic fluids.
The directed motion present in active systems results in a stronger separation for particle speeds and longer correlation lengths than would be seen in traditional systems when considering phase identity near the critical point.
Ha \textit{et al.} were able to successfully characterize particle phase of a Lennard-Jones fluid with high accuracy using a convolutional neural network and only three structural features,\cite{Ha2018} but we find that our model performance steeply drops off near the critical point if we do not include at least one of the kinematic features mentioned above.

We have demonstrated that the local structure plays an important role in determining the phase behavior of active systems.
Our graph representations of the system possess unique characteristics specific to their region of the phase diagram---which can be learned using a general graph neural network framework with attention. 
This matches the results from Swanson \textit{et al.} and Ha \textit{et al.} who included structure via a message-passing network and convolutional neural network, respectively, to characterize amorphous materials.\cite{Ha2018, Swanson2020DeepMaterials}

We believe machine learning can be used for more challenging classification problems. 
It would be straightforward to extend our framework to also distinguish between the hexatic crystalline phase and the disordered dense phase to produce a more complete phase diagram. 
Our model is already capable of learning the importance of the third shell average Voronoi volumes, which act as a surrogate for the third peak in the radial distribution function.
This peak provides a way to distinguish between liquid and solid phases.
We also feel a more specific model could be created to directly predict which region of the phase diagram the system is in by performing classification at the graph-level instead of the node-level (as was done in this work).
A graph-level classifier can then be readily generalized using an unsupervised learning scheme, where the model is learning distinctions between the present phases.

The learning architecture used here should also readily generalize to active systems with thermal noise, polydispersity, or higher dimensionality.
These deviations from the problem focused on in this work would result in different distributions for feature values, but should still maintain similar relationships between features.
The graph network can be further extended to include edge features, which would allow for more complicated or varied interparticle interactions and should prove to be a useful tool in the characterization of other amorphous systems.

\section*{Data Availability}
The data that support the findings of this study are available from the corresponding author upon reasonable request.

\acknowledgements
A.R.D. would like to thank Yisong Yue for thoughtful discussions pertaining to graph neural networks.
J.F.B. acknowledges support by the National Science
Foundation under Grant No. CBET-1803662. We gratefully acknowledge the support of the NVIDIA Corporation for the donation of the Titan V GPU used to carry
out this work.

\appendix
\section{\label{app:feature_corr} Feature Correlation and Importance}

\begin{figure}[tb]
    \centering
    \includegraphics[width=0.49\textwidth]{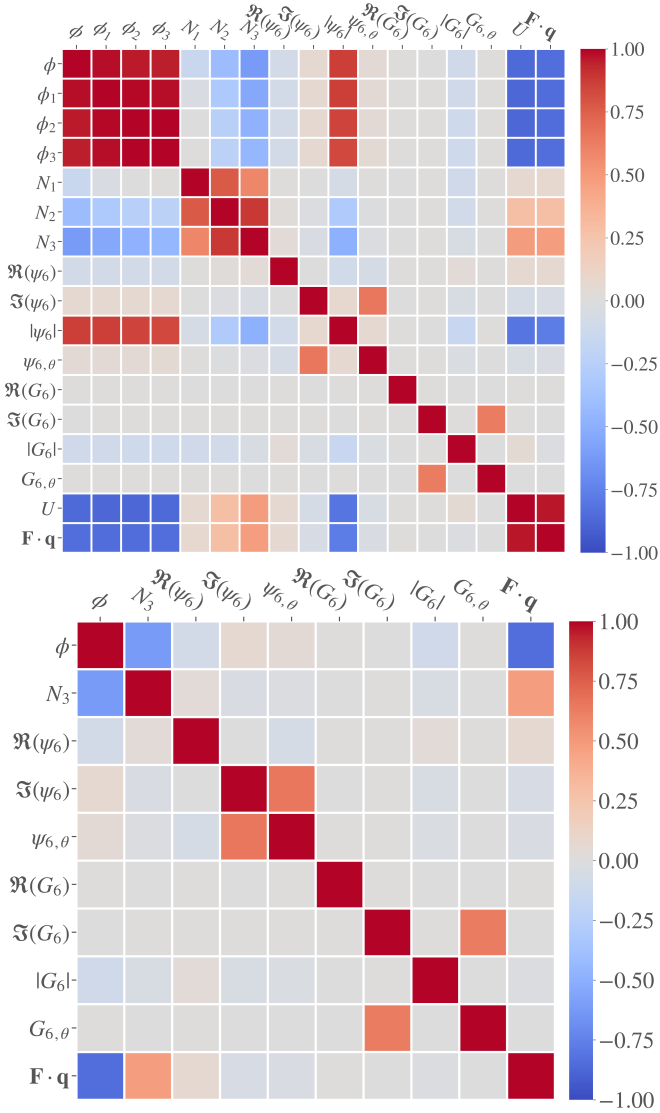}
    \caption{The correlation matrix for the (\textbf{top}) full and (\textbf{bottom}) reduced feature sets.
    Strong positively (red) and negatively (blue) correlated features are removed in the reduced feature set.
    }
    \label{fig:correlations}
\end{figure}

The correlation matrix for our full initial feature set is presented at the top of Fig.~\ref{fig:correlations}.
In the figure we have used a shorthand notation where $\phi$ is the Voronoi volume fraction, $\phi_{i}$ is the Voronoi volume fraction averaged over the $i^{th}$ shell neighbors, $N_{i}$ is the number of neighbors in shell $i$, $U$ is the particle speed, $\mathbf{F}\cdot\mathbf{q}$ is the force-orientation correlation, $\psi_6$ is the hexatic order parameter, and $G_{6}$ is the translational order parameter.
The hexatic and translational order parameters are broken into their real part, imaginary part, magnitude, and angular components represented by $\Re(\cdot)$, $\Im(\cdot)$, $|\cdot|$, and $(\cdot)_{6,\theta}$, respectively.
Features were dropped in order of the strength of the measured collinearity with other features.
When considering a pair of collinear features, the feature that contributes the least to the total importance is removed.

\begin{figure}[tb]
    \centering
    \includegraphics[width=0.49\textwidth]{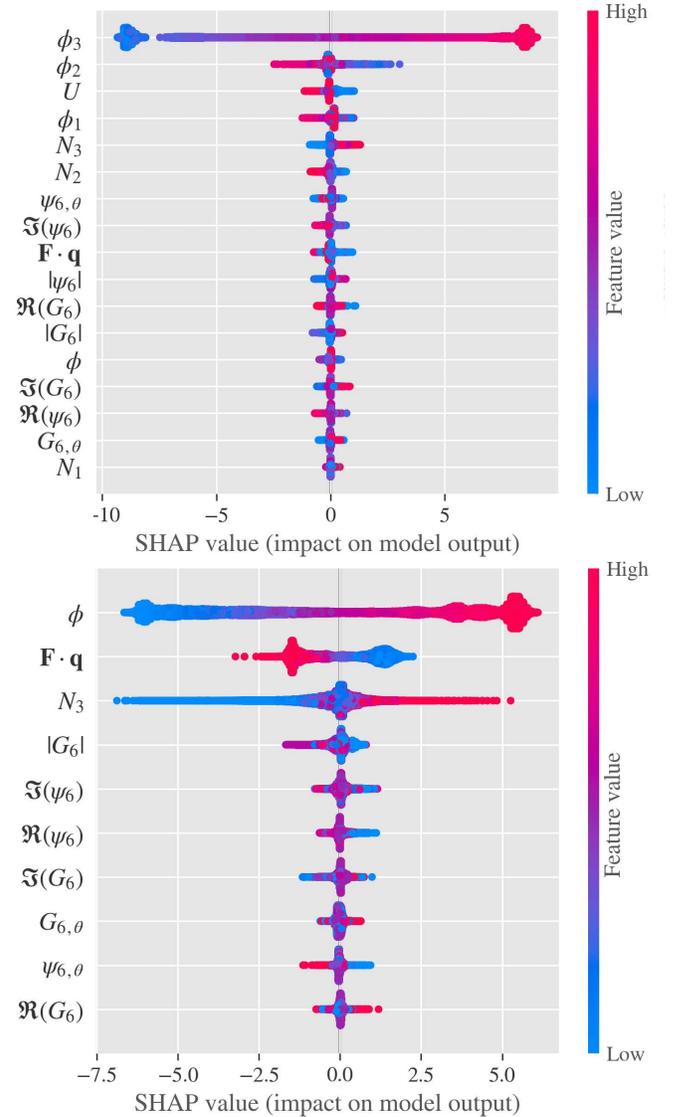}
    \caption{Feature importance for the (\textbf{top}) full and (\textbf{bottom}) reduced feature sets computed using SHAP. 
    The color corresponds to the magnitude of a given feature.
    The SHAP value presents how important a feature is at predicting the positive class.}
    \label{fig:shap}
\end{figure}

We use a simple boosted random forest to compute feature importance. 
Our random forest classifier is made up of 1000 estimators, with a max decision tree depth of 8, and trained for 30 epochs with early stopping.
Our boosted random forest is implemented in XGBoost.
This classifier is then used to compute the SHAP feature importance (see Fig.~\ref{fig:shap}).\cite{Lundberg2017APredictions}
The color in Fig.~\ref{fig:shap} indicates the value of the feature in the line, and the actual SHAP value indicates how important a feature value was for predicting the positive (dense) case.
As an example, from Fig.~\ref{fig:shap}(top) we see that $\phi_{3}$ is the most indicative feature, and large values of this feature strongly indicate that the particle is dense, whereas very low values indicate that the particle in question is likely dilute.
The SHAP analysis for the full feature set is not very insightful due to the presence of strong collinearity, but it can still be used to determine which feature to drop from a pair of highly collinear features.
After removing a feature the importance is recalculated as this can change as the feature set changes.
The final correlation matrix for the features used in this work is shown at the bottom of Fig.~\ref{fig:correlations}, and the final feature SHAP values are shown in the bottom of Fig.~\ref{fig:shap}.
There is greater diversity in the SHAP values obtained, and now the volume fraction as the most important feature, which is in line with our physical intuition. 

\section{\label{app:model_details} Model and Training Details}

\begin{table}[tb]
\caption{\label{table:model} The specific model architecture of the trained deep neural network used for the results presented in this work.}
\begin{ruledtabular}
\begin{tabular}{ccccc}
Layer&Size&
\multicolumn{1}{c}{\textrm{Activation}}&
\multicolumn{1}{c}{\textrm{Batch Norm}}&
\multicolumn{1}{c}{\textrm{Dropout}}\\
\hline
1 & 128 & ReLU & -- & --\\
2 & 128 & LeakyReLU\footnote{\label{foot:leakyrelu}LeakyReLU activation function has negative slope  $\alpha=0.1$.}
                    & True & 0.69 \\
3 & 128 & LeakyReLU\textsuperscript{\ref{foot:leakyrelu}}
                    & True & 0.35 \\
4 & 64 & LeakyReLU\textsuperscript{\ref{foot:leakyrelu}}
                    & -- & 0.75 \\
5 & 2 & SoftMax & -- & -- \\
\end{tabular}
\end{ruledtabular}
\end{table}
\begin{table}[tb]
\caption{\label{table:gnn} The architecture of the graph network portion of our model.}
\begin{ruledtabular}
\begin{tabular}{cccc}
Layer&Size&
\multicolumn{1}{c}{\textrm{Activation}}&
\multicolumn{1}{c}{\textrm{Attention Heads}}\\
\hline
1 & 8 & LeakyReLU\footnote{\label{foot:gnn_relu}LeakyReLU activation function has negative slope  $\alpha=0.2$.}
                & 2\\
2 & 8 & LeakyReLU\textsuperscript{\ref{foot:gnn_relu}}
                    & 2 \\
3 & 8 & LeakyReLU\textsuperscript{\ref{foot:gnn_relu}}
                    & 2 \\
4\footnote{This is a fully-connected layer used to get the final prediction.} & 2 & SoftMax & -- \\
\end{tabular}
\end{ruledtabular}
\end{table}

The trained DNN used in this work is 5 layers with batch normalization and dropout on some of the layers for regularization.
The number of layers in the network, size of each layer, batch normalization, and dropout values were determined from 1,500 rounds of hyperparameter optimization with the Hyperopt package.\cite{Bergstra2013MakingArchitectures}
The hyperparameter optimization was performed in three stages, each of which was 500 rounds.
We first optimize the learning rate to speed up future training as much as possible.
The optimal learning rate $lr=3\times10^{-3}$ was used for the remaining optimization rounds with a batch size of 32.
The next round of optimization focuses on the number of neurons in the network, the number of layers, and the activation function used for the layer (between ReLU and LeakyReLU).
The final optimization round is focused on regularization and tunes the batch normalization and dropout for each layer in the network.
Our final chosen parameters are presented in Table~\ref{table:model}.

The GNN model architecture used in this work was explored manually.
The graph network is intentionally kept small as this was shown by Veli\v{c}kovi\'{c} \textit{et al.} to be effective at transductive learning.\cite{Velickovic2018GraphNetworks}
The parameters of our GNN are shown in Table~\ref{table:gnn}.
Each layer in the network is a GAT convolution layer except for the last one, which is a fully-connected layer to give the outputs.

\section*{References}

%
\end{document}